\documentclass[prb,twocolumn,showpacs]{revtex4}     
\usepackage{epsfig,amsmath} 

\begin{document}

\title{Compressibility of Interacting Electrons in Bilayer Graphene} 

\author{Xin-Zhong Yan$^{1}$ and C. S. Ting$^2$}
\affiliation{$^{1}$Institute of Physics, Chinese Academy of Sciences, P.O. Box 603, 
Beijing 100190, China\\
$^{2}$Texas Center for Superconductivity, University of Houston, Houston, Texas 77204, USA}
 
\date{\today}
 
\begin{abstract}
Using the renormalized-ring-diagram approximation, we study the compressibility of the interacting electrons in bilayer graphene. The chemical potential and the compressibility of the electrons can be significantly altered by an energy gap (tunable by external gate voltages) between the valence and conduction bands.  For zero gap and a typical finite gap in the experiments, we show both systems are stable. \end{abstract}

\pacs{73.22.Pr,81.05.ue,51.35.+a} 

\maketitle

Bilayer graphene has attracted considerable attention because of its promised application in electronic devices.\cite{Ohta, Henriksen, Young, Martin, McCann1, Min, McCann2, Koshino, Nilsson1, Barlas1, Nilsson2, Wang, Hwang, Barlas2, Borghi1, Nandkishore, Borghi2, Kusminskiy1, Borghi3} In contrast to the Dirac fermions in monolayer graphene, the energy bands of the free electrons in bilayer graphene are hyperbolic and gapless between the valence and conduction bands.\cite{McCann1} Most importantly, an energy gap opening between the valence and conduction bands can be generated and controlled by external gate voltages. At low carrier doping, because the energy-momentum dispersion relation around the Fermi level is relatively flat, the Coulomb effect is expected to be significant in the interacting electron system of bilayer graphene. The Coulomb effect has been studied by a number of theoretical works using the Hartree-Fock (HF) and random-phase approximations.\cite{Nilsson2, Wang, Hwang, Barlas2, Borghi1, Nandkishore, Borghi2, Kusminskiy1, Borghi3} One of the thermodynamic quantities directly reflecting the Coulomb effect is the electronic compressibility. Recently, this quantity has been measured by experiments.\cite{Henriksen,Young,Martin} For timely coordinating with the experimental measurements, it is necessary to theoretically study the combined effect of Coulomb interactions and the gap opening in the compressibility with a more realistic model.

In this work, using the renormalized-ring-diagram approximation (RRDA),\cite{Yan,Yan1} we study the Coulomb effect in the compressibility of the interacting electrons in bilayer graphene. The chemical potential and its derivative with respect to the carrier density are calculated for systems of zero gap and a typical gap in the existing experiments. Though a finite-gapped system is more perturbable by the Coulomb interaction than the zero-gapped one at low carrier doping and low temperature, we show both systems studied here are stable. 

The atomic structure of bilayer graphene is shown in Fig. 1. The two sublattices in each layer are denoted by A (white) and B (black) atoms, respectively. The interlayer distance is $c = 3.34$ \AA~ $\approx 1.4 a$ where $a \approx 2.4$ \AA~is the lattice constant of monolayer graphene (the distance between two nearest A atoms). The energy of electron hopping between the nearest-neighbor (nn) carbon atoms in each layer is $t \approx 2.82$ eV,\cite{Bostwick} while the interlayer nn hopping is $t_1 \approx 0.39$ eV.\cite{Misu} 

The Hamiltonian describing the electrons is given by
\begin{eqnarray}
H=\sum_{j}\epsilon_jn_j-\sum_{ij\sigma}t_{ij}c^{\dagger}_{i\sigma}c_{j\sigma}+\frac{1}{2}\sum_{ij}\delta n_iv_{ij}\delta n_j \label {hm}
\end{eqnarray}
where $\epsilon_j = \pm \Delta$ for top (back) layer, $n_j$ is the electron density at site $j$, $c^{\dagger}_{i\sigma}$ creates an electron at site $i$ with spin $\sigma$, $\delta n_j=n_j-n$ with $n$ the average occupation number of electrons per site (which is also the charge number of the neutralizing background), $v_{ij}$ is the Coulomb interaction between electrons at sites $i$ and $j$.  Here $\Delta$ is the energy gap parameter, which is a consequence of the potential difference between the top and back gates (attached to the top and back layers, respectively). The model is restricted to nn hopping within the same layer and between the adjacent sites on top and back layers as shown in Fig. 1. The Coulomb interaction is given by $v_{ij} = e^2/\epsilon r_{ij}$ for $r_{ij} \ne 0$ with $\epsilon \sim 4$ the dielectric constant of high frequency limit of the system; $r_{ij}$ here is the distance between two electrons at sites $i$ and $j$ in the bilayer system. For onsite Coulomb interaction $U = v_{jj}$, we take $U$ in our calculation as $U = 2v_{AB}$ where AB denotes the nn sites in the same layer. The dimensionless Coulomb coupling constant is given by $e^2/\epsilon at \approx 0.5$. As described by Eq. (\ref{hm}), we here consider only the charge-charge interactions. The onsite interaction may lead to the antiferromagnetic correlation. We here neglect it since it is not important in graphene. 
 
\begin{figure} 
\centerline{\epsfig{file=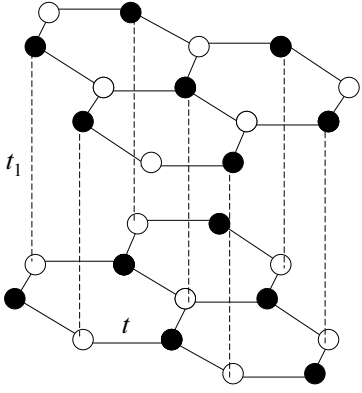,width=5.5 cm}}
\caption{Structure of Bernal stacking bilayer graphene. The energies of intra- and interlayer electron hopping between the A (white) and B (black) atoms are given by $t$ and $t_1$, respectively. } 
\end{figure} 

By defining the doped electron concentration $\delta$ per carbon atom, we have $n = 1+\delta$. Under the transformation $\delta \to -\delta$ and $c_{j,\sigma} \leftrightarrow \pm c^{\dagger}_{j,\sigma}$ for electrons at A (B) site, $H$ is unchanged. Therefore our system satisfies the particle-hole symmetry. Furthermore, $K = H -\mu(\hat N - N_0)$ (with $\hat N$ the operator of the total number of electrons and $N_0$ the total number of lattice sites and thereby $N - N_0$ as the total number of doped electrons) is also unchanged under the above transformation and $\mu \to -\mu$. Thus, the chemical potential $\mu$ is an odd function of $\delta$.

To proceed, we begin with the Green's function $G$ defined as 
\begin{eqnarray}
G(i,j,\tau-\tau') = -\langle T_{\tau}C_{i\sigma}(\tau)C^{\dagger}_{j\sigma}(\tau')\rangle \label {gf}
\end{eqnarray}
where $C^{\dagger}_{j\sigma} = (c^{\dagger}_{a1j\sigma},c^{\dagger}_{b1j\sigma},c^{\dagger}_{a2j\sigma },c^{\dagger}_{b2j\sigma})$ with $c^{\dagger}_{a(b)\mu j\sigma}$ creating an electron of spin $\sigma$ at site A (B) of $\mu$th (= 1,2 respectively for top and back) layer of $j$th unit cell. In the momentum-frequency space, $G$ is expressed as
\begin{eqnarray}
G(k,i\omega_{\ell}) = [i\omega_{\ell} -\xi_k-\Sigma(k,i\omega_{\ell})]^{-1} \label {Gk}
\end{eqnarray}
with
\begin{eqnarray}
\xi_k = \begin{pmatrix}
-\mu& \epsilon_k&0&0\\
\epsilon_k^{\ast}&-\mu&t_1&0\\
0&t_1&-\mu&\epsilon_k\\
0&0&\epsilon_k^{\ast}&-\mu\\
\end{pmatrix}
\end{eqnarray}
where $\omega_{\ell} = (2\ell+1)\pi T$ is the fermionic Matsubara frequency with $\ell$ as integer number and $T$ the temperature, $\epsilon_k = t_0[1+\exp(-ik_x)+\exp(-ik_y)]$, $\mu$ is the chemical potential, and $\Sigma(k,i\omega_{\ell})$ is the self-energy. Under RRDA, $\Sigma$ is geven by the functional derivative of the `free energy' functional $\Phi$ with respective to $G$ shown diagrammatically in Fig. 2,
\begin{eqnarray}
\Sigma =\delta \Phi/\delta G. \label {se}
\end{eqnarray}
In Fig. 2, each bubble is composed by two renormalized Green's function $G$'s. The last terms in the expression of $\Phi$ and $\Sigma$ stem from the additional term in writing the product of two density operators appeared in the interaction term of Eq. (\ref{hm}) in normal order of electron operators: $c^{\dagger}_{i\sigma}c_{i\sigma}c^{\dagger}_{j\sigma}c_{j\sigma}=c^{\dagger}_{i\sigma}c^{\dagger}_{j\sigma}c_{j\sigma}c_{i\sigma} +\delta_{ij}c^{\dagger}_{j\sigma}c_{j\sigma}$. Note that Eq. (\ref{se}) is a $4\times 4$ matrix equation. In terms of $G$, the elements of $\Sigma$ are expressed as 
\begin{eqnarray}
\Sigma_{\mu\nu}(k,i\omega_{\ell}) &=& -\frac{T}{M}\sum_{q,m} G_{\mu\nu}(k-q,i\omega_{\ell}-i\nu_m)W_{\mu\nu}(q,i\nu_m)\nonumber\\
&& +\delta_{\mu\nu}U/2 \nonumber
\end{eqnarray}
where $M = N_0/4$ is total number of unit cells in the system, $\nu_m$ is the bosonic Matsubara frequency, and $W_{\mu\nu}(q,i\nu_m)$ is an effective interaction. The matrix form of $W$ is given by
\begin{eqnarray}
W(q,i\nu_m) = [1-v(q)\chi(q,i\nu_m)]^{-1}v(q) \nonumber
\end{eqnarray}
with 
\begin{eqnarray}
\chi_{\mu\nu}(q,i\nu_m)=\frac{2T}{M}\sum_{k,\ell} G_{\mu\nu}(k,i\omega_{\ell})G_{\nu\mu}(k-q,i\omega_{\ell}-i\nu_m) \nonumber
\end{eqnarray}
and $v(q)$ is the Fourier component (matrix form) of the Coulomb interaction. In the present lattice model, $v(q)$ cannot be expressed explicitely. A similar calculation of $v(q)$ for the single layer graphene is given in Ref. \onlinecite{Yan1}. The chemical potential $\mu$ is determined by
\begin{eqnarray}
n = \frac{2T}{N_0}\sum_{k\ell}{\rm Tr}G(k,i\omega_{\ell})\exp(i\omega_{\ell}\eta), \label {chm}
\end{eqnarray}
where $\eta$ is an infinitesimal small positive constant. The Green's function so determined satisfies the microscopic conservation laws.\cite{Baym}

\begin{figure} 
\centerline{\epsfig{file=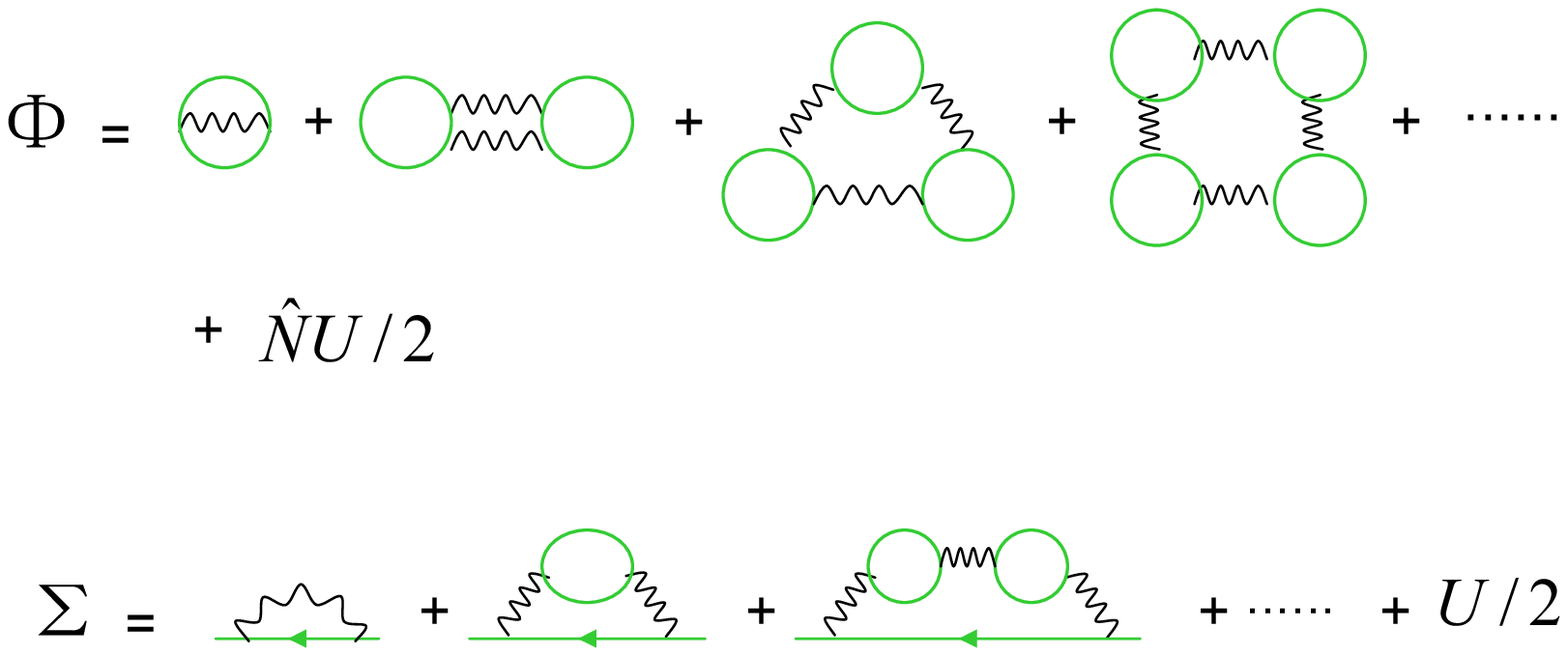,width=9. cm}}
\caption{(color online) Diagrammatic expressions for `free energy' functional $\Phi$ and self-energy $\Sigma$. The line represents the Green's function and the wavy line is the Coulomb interaction. The last terms stem from the additional term in writing the product of two density operators in normal order of electronic operators. $U$ is the onsite Coulomb interaction.}
\end{figure} 

The behavior of the chemical potential is closely related to the compressibility $\kappa$ of the doped carrier system. It is defined as
\begin{equation}
\kappa = \frac{1}{n^2}(\frac{\partial n}{\partial\mu})_T.   \label{cmp}			
\end{equation}
Its inverse $(\partial\mu/\partial n)_T$ can be calculated by numerical derivative of $\mu$ determined by Eq. (\ref{chm}) with respect to $n$. Clearly, when $(\partial\mu/\partial n)_T$ goes to zero, $\kappa$ becomes infinity, implying an inhomogeneity tendency in the system.

\begin{figure} 
\centerline{\epsfig{file=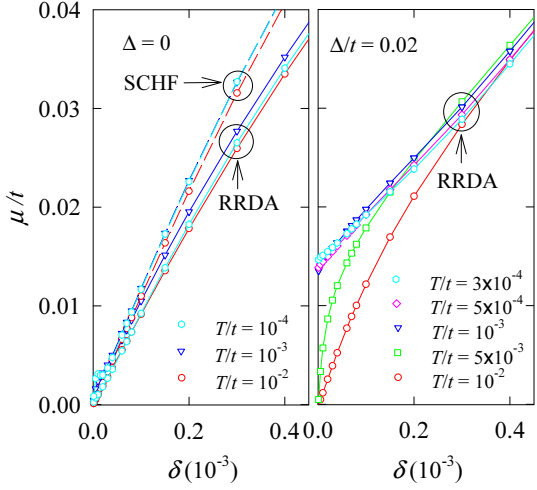,width=7.5 cm}}
\caption{(color online) Chemical potential $\mu$ as function of electron doping concentration $\delta$ at various temperatures calculated by SCHF and RRDA for zero-gapped systems (left panel) and by RRDA for systems of $\Delta/t = 0.02$ (right panel).  }  
\end{figure} 

Figure 3 shows the chemical potential $\mu$ as a function of $\delta$ for the systems of $\Delta = 0$ (left panel) and $\Delta/t =0.02$ (right panel) at various temperatures. The results by RRDA are compared with that by the self-consistent Hartree-Fock approximation (SCHF) (which is obtained from $\Phi$ by neglecting all other ring terms except the first one) for the zero-gapped system. The chemical potential $\mu$ for zero gap case goes to zero as $\delta \to 0$ because of the particle-hole symmetry. The overall behaviors of $\mu$ given by RRDA for system of $\Delta = 0$ at various temperatures are smooth and seem not much different from each other. Physically, at low doping, the electrons in the valence band can be thermally excited to high levels in the conduction band without significant change in the chemical potential because the density of states is symmetric about the touch point of the valence and conduction bands. The same feature is seen by SCHF. However, in contrast to the smooth behavior given by RRDA, the curve corresponding to lower temperature given by SCHF is twisty in a region close to zero doping. This behavior will result in a singularity in the compressibility. On the other hand, for the gapped system, the chemical potentials are indistinguishable only at very low temperatures. At low temperature, $\mu$ is finite at $\delta = 0$. For free electron system, $\mu$ approaches $\Delta$ as $\delta \to 0^+$ and $T \to 0$. The limit here is less than $\Delta$ because of inter-electronic Coulomb interactions. Since the limits $\delta \to 0^{\pm}$ are different, there is a discontinuity in $\mu$ at $\delta = 0$ from hole side to electron side. With increasing the temperature, $\mu$ is suppressed due to thermal excitations between the valence and conduction bands. The thermal effect is significant at lower carrier doping since where the Fermi energy is lower. At high temperature, $\mu$ at $\delta = 0^+$ is bound from below, $\mu \ge 0$, because of the particle-hole symmetry.

\begin{figure} 
\centerline{\epsfig{file=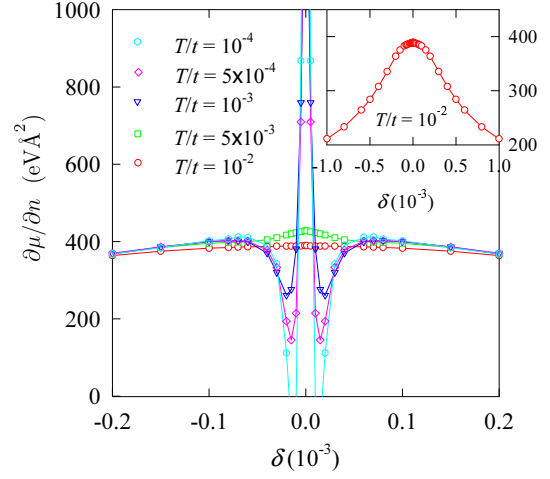,width=7.5 cm}}
\caption{(color online) $(\partial\mu/\partial n)_T$ as function of doping concentration $\delta$ at various temperatures obtained by SCHF for $\Delta = 0$. The inset shows the result at $T/t = 10^{-2}$ for a wide range of doping. } 
\end{figure} 

For illustrating the RRDA result, we firstly show in Fig. 4 the one by SCHF for $(\partial\mu/\partial n)_T$ as a function of $\delta$ for $\Delta = 0$ at various temperatures. At low temperature, there is a peak at $\delta = 0$. This is because the chemical potential at finite doping is suppressed by the Coulomb interactions but not changed at $\delta = 0$ due to the particle-hole symmetry. At low but finite doping close to $\delta = 0$, there is a sharp decrease in $(\partial\mu/\partial n)_T$ as $T \to 0$. This stems from the twisty behavior in $\mu$. At $T = 10^{-4}$, $(\partial\mu/\partial n)_T$ becomes negative for electron/hole concentration within a small region. At $T = 0$, the result $(\partial\mu/\partial n)_T < 0$ at low doping has been predicted by Kusminskiy {\it et al.} using perturbative HF approach on a continuum model.\cite{Kusminskiy2} The compressibility is singular at the doping where $(\partial\mu/\partial n)_T = 0$. This is a strong Coulomb effect at low doping where the Fermi level of the electrons is low. This is similar to the prediction of the Wigner crystallization in the two-dimension electron gas at low density. The inset shows the result at $T/t = 10^{-2}$ for a wide range of doping. All curves approach the same behavior at large doping as the one at $T/t = 10^{-2}$. 

\begin{figure} 
\centerline{\epsfig{file=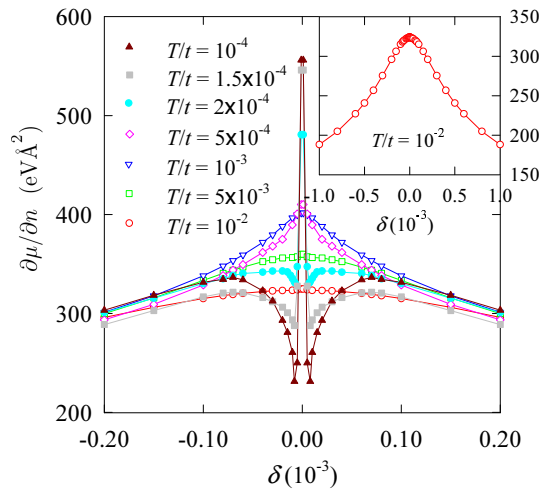,height=7.cm}}
\caption{(color online) $(\partial\mu/\partial n)_T$ as function of doping concentration $\delta$ at various temperatures obtained by RRDA for $\Delta = 0$. The inset shows the result at $T/t = 10^{-2}$ for a wide range of doping. } 
\end{figure} 

The differences between Ref. \onlinecite{Kusminskiy2} and the SCHF calculation here are in following aspects. (1) Ref. \onlinecite{Kusminskiy2} uses the perturbative HF in which the Green's function not renormalized as done in SCHF. (2) The lattice model here is different from the continuum model adopted in Ref. \onlinecite{Kusminskiy2}. In the continuum model, the intralayer and interlayer interactions are given by $v(q) = 2\pi e^2/q$ and $2\pi e^2\exp(-cq)/q$, respectively, with a cutoff $q_c$. Because of this cutoff, the interactions in real space are not the Coulomb form $e^2/r_{ij}$ that is used here. The compressibility $\kappa$ should sensitively depend on the interactions. (3) By the lattice model, the particle-hole symmetry is satisfied. On the other hand, there is no such a symmetry in the continuum model. (4) The present SCHF calculation is for finite temperature, different from Ref. \onlinecite{Kusminskiy2} that the calculation is performed for $T = 0$. 

In general, both HF and SCHF for the systems of long-range Coulomb interactions over count the strong coupling effects. More terms arising from the interactions should be included in the `free energy' functional. By RRDA, all the ring terms that are the most long-wavelength divergent terms are summed over in $\Phi$. As a result, the effective interactions between electrons are weakened from the bare Coulomb one. Shown in Fig. 5 is $(\partial\mu/\partial n)_T$ obtained by RRDA as a function of $\delta$ for the zero gap system at various temperatures. The results given in Fig. 5 are quite different from that in Fig. 4. At temperatures down to $T/t = 5\times 10^{-4}$, $(\partial\mu/\partial n)_T$ monotonically decreases with the electron/hole doping concentration. At $T/t \le 2\times 10^{-4}$, the Coulomb suppression shows up nearby the central peak similarly as in Fig. 4. But the deviation from the monotonically decreasing behavior is much smaller than that given by SCHF. From the low temperature behaviors of $(\partial\mu/\partial n)_T$, we deduce that it is positive at $T = 0$ and the system is stable.

The results for a gapped system of typical $\Delta/t = 0.02$ as in experiments are depicted in Fig. 6. By comparing to the zero gap case, the behavior of $(\partial\mu/\partial n)_T$ is delicate. At temperature higher than the gap, because there are significant excitations between the valence and conduction bands, both zero and finite-gapped systems are not different so much. However, at low temperature, $(\partial\mu/\partial n)_T$ is suppressed much in a wide doping range around $\delta = 0$. This is because of the special conduction band structure of the gapped system. For $\Delta = 0$, the band bottom is parabolic. At finite $\Delta$, however, the location of the minimum of the conduction band (for electron doping) is a ring surrounding the Dirac point. In a region encircling the ring, the energy band is nearly flat. Therefore, at low doping, the energy of electrons is approximately dispersionless. For such a dispersionless band, the free electron system is more perturbable under the Coulomb interactions. Even though the quantity $(\partial\mu/\partial n)_T$ is considerably suppressed, by carefully looking at its behavior at low $T$, it is positive in the limit $T \to 0$.

The $(\partial\mu/\partial n)_T \le 0$ instability is not observed in recent experiments on the compressibility.\cite{Henriksen,Young,Martin} In the experiment of Ref. \onlinecite{Henriksen}, the quantity $(\partial\mu/\partial n)_T$ is observed as monotonically deceasing in both sides of the electron and hole doping with a central peak around the neutral point. In Ref. \onlinecite{Martin}, the observation at zero magnetic field shows that the central peak in $(\partial\mu/\partial n)_T$ is very sharp and the quantity then varies nearly flat within a wide range of electron/hole doping. While in the experiment of Ref. \onlinecite{Young}, the structure of central peak-dip-hump in $(\partial\mu/\partial n)_T$ as function of doping is observed (at zero magnetic field). Though the experiments do not quantitatively agree with each other, the common fact is $(\partial\mu/\partial n)_T > 0$. As for the dip or the flat region nearby the central peak in $(\partial\mu/\partial n)_T$, from the opinion of the present calculation, they may stem from the Coulomb suppression effect as studied here. However, the calculation given here is not in quantitatively agreement with the experimental results. So far there is no theory can quantitatively explain the experiments. To solve the problem, besides the long range Coulomb interaction, other kinds of effects may need to be taken into account.

\begin{figure} 
\centerline{\epsfig{file=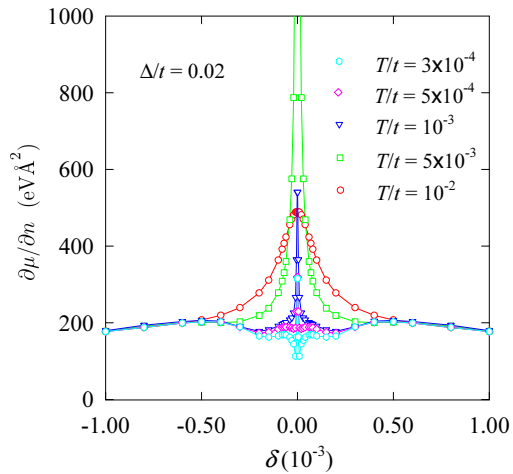,height=7.cm}}
\caption{(color online) $(\partial\mu/\partial n)_T$ as function of doping concentration $\delta$ at various temperatures obtained by RRDA for $\Delta = 0.02$ gapped system. } 
\end{figure} 

In summary, we have investigated the compressibility of the interacting electrons in bilayer graphene using RRDA. For the zero gap system, with comparing to the result by SCHF, the compressibility given by RRDA is positive. The homogeneous system with zero gap is therefore stable. The system with a finite-gap is more perturbable by the Coulomb interaction than the one of zero gap. For the typical gapped system as investigated in experiments, however, we find that the system is stable as well.  

This work was supported by the Robert A. Welch Foundation under Grant No. E-1146, the TCSUH, the National Basic Research 973 Program of China under Grant No. 2011CB932700, NSFC under Grants No. 10774171 and No. 10834011, and financial support from the Chinese Academy of Sciences for advanced research.

\end{document}